%% file: PUB_topLSP1lepton.tex
\documentclass[prl,superscriptaddress,twocolumn,showpacs,preprintnumbers,amsmath,amssymb]{revtex4-1}

\usepackage{subfigure}
\usepackage{epsfig}
\usepackage{dcolumn}
\usepackage{bm}
\usepackage{graphicx}
\usepackage{endnotes}
\usepackage{multirow}
\usepackage[hyperindex,breaklinks]{hyperref}

\usepackage{xspace}
\usepackage[sort&compress]{natbib}

\usepackage{preprintcover}
\usepackage{atlasphysics}
\usepackage{draftcopy}

\newcommand{\hideit}[1]{{ }}

\newcommand{\alpgen}{{\tt ALPGEN}}
\newcommand{\alpgenv}{{\tt ALPGEN~2.14}}
\newcommand{\powheg}{{\tt PowHeg}}
\newcommand{\powhegv}{{\tt PowHeg HVQ patch 4}}

\newcommand{\herwig}{{\tt HERWIG}}
\newcommand{\herwigv}{{\tt HERWIG~6.520}}

\newcommand{\herwigppv}{{\tt Herwig++~2.5.2}}
\newcommand{\pythia}{{\tt PYTHIA}}
\newcommand{\pythiav}{{\tt PYTHIA~6.425}}
\newcommand{\mcatnlo}{{\tt MC@NLO}}
\newcommand{\mcatnlov}{{\tt MC@NLO~4.01}}
\newcommand{\fewz}{{\tt FEWZ}}
\newcommand{\mcfm}{{\tt MCFM}}

\newcommand{\jimmyv}{{\tt JIMMY~4.31}}
\newcommand{\cteqVER}[1]{{\tt CTEQ#1}}

\newcommand{\madgraph}{{\tt MADGRAPH}}
\newcommand{\madgraphv}{{\tt MADGRAPH~5}}

\newcommand{\acermcv}{{\tt AcerMC~3.7}}

\def\CLs{\ensuremath{\mathrm{CL}_s}}

\newcommand{\oneLepTitle}{Search for Direct Top Squark Pair Production in Final
States with One Isolated Lepton, Jets, and Missing Transverse Momentum in $\sqrt{s}=7$~\TeV\ $pp$ Collisions using $4.7$~\ifb\ of ATLAS Data}

\PreprintCoverPaperTitle{\oneLepTitle}
\PreprintIdNumber{CERN-PH-EP-2012-200} 
\PreprintCoverAbstract{  A search is presented for direct top squark pair production in final
  states with one isolated electron or muon, jets, and missing
  transverse momentum in proton-proton collisions at 
  $\sqrt{s} = 7\,\TeV$.
  The measurement is based on 4.7~\ifb\ of data
  collected with the ATLAS detector at the LHC.  Each top squark is
  assumed to decay to a top quark and the lightest supersymmetric
  particle (LSP). The data are found to be consistent with standard
  model expectations. Top squark masses between $230$\,\GeV\ and $440$\,\GeV\ are
  excluded with 95\% confidence for massless LSPs, and top squark masses
  around $400$\,\GeV\ are excluded for LSP masses up to $125$\,\GeV.}  
\PreprintJournalName{Physical Review Letters}

\newcommand{\tone}   {\tilde{t}^{}_{1}}

\def\bjet{$b$-jet\xspace}

\def\mt{\ensuremath{m_{\mathrm{T}}}\xspace}
\def\metsig{\ensuremath{\mathrm{MetSig}}\xspace}

\def\Wjets{$W$+jets\xspace}
\def\Zjets{$Z$+jets\xspace}

\renewcommand{\ttbar} {\ensuremath{t\bar{t}}\xspace}

\renewcommand{\met} {\ensuremath{E_{\mathrm{T}}^{\mathrm{miss}}}\xspace}
\renewcommand{\pt} {\ensuremath{p_\mathrm{T}}\xspace}

\def\met{\ensuremath{E_{\mathrm{T}}^{\mathrm{miss}}}}

\def\metsig{\ensuremath{E_\mathrm{T}^{\mathrm{miss}}/\sqrt{ \HT }}}

\begin{document}
\title{\oneLepTitle}
\author{The ATLAS Collaboration}

\begin{abstract} 
  A search is presented for direct top squark pair production in final
  states with one isolated electron or muon, jets, and missing
  transverse momentum in proton-proton collisions at 
  $\sqrt{s} = 7\,\TeV$.
  The measurement is based on 4.7~\ifb\ of data
  collected with the ATLAS detector at the LHC.  Each top squark is
  assumed to decay to a top quark and the lightest supersymmetric
  particle (LSP). The data are found to be consistent with standard
  model expectations. Top squark masses between $230$\,\GeV\ and $440$\,\GeV\ are
  excluded with 95\% confidence for massless LSPs, and top squark masses
  around $400$\,\GeV\ are excluded for LSP masses up to $125$\,\GeV.
\end{abstract}

\pacs{14.80.Ly}
\maketitle
\normalsize

Weak scale supersymmetry
(SUSY)~\cite{Miyazawa:1966,Ramond:1971gb,Golfand:1971iw,Neveu:1971rx,Neveu:1971iv,Gervais:1971ji,Volkov:1973ix,Wess:1973kz,Wess:1974tw}
is an extension to the standard model (SM) that provides a solution to
the hierarchy problem by introducing supersymmetric partners of all SM
particles. In the framework of a generic $R$-parity conserving minimal
supersymmetric extension of the SM
(MSSM)~\cite{Fayet:1976et,Fayet:1977yc,Farrar:1978xj,Fayet:1979sa,Dimopoulos:1981zb},
SUSY particles are produced in pairs, and the lightest supersymmetric
particle (LSP) is stable and can be a dark matter candidate. In a large
variety of models, the LSP is the lightest neutralino,
$\tilde{\chi}^{0}_{1}$, which only interacts weakly and thus escapes detection.

Light top squarks (stop) are suggested by naturalness
arguments~\cite{Barbieri198863,de_Carlos:1993yy}.
Searches for direct stop pair production have been previously 
reported by the CDF and D0 experiments~\cite{PhysRevLett.104.251801,D0_stopSearch}. 
Searches for stops via $\gluino \gluino$ production have been reported by the
ATLAS~\cite{ATLAS:2012ah,PhysRevLett.108.241802,Aad:2012hm} and
CMS~\cite{cmsBjetsMET,cmsSSlepBjet} collaborations.
In this Letter, one stop mass eigenstate ($\tone$) is assumed to be
significantly lighter than the other squarks.  A search is presented
for directly pair-produced stops, which are each assumed to decay to a
top quark and the LSP. The signature for such a signal is
characterized by a top quark pair ($t\overline{t}$) produced in
association with possibly large missing transverse momentum, the magnitude of
which is referred to as \met, from the undetected LSPs.
The analysis targets final states where one top quark decays hadronically and the other semileptonically.
%One top quark is assumed to decay hadronically and the other semileptonically.

The ATLAS detector~\cite{Aad:2008zzm} has a solenoid, surrounding the
inner tracking detector (ID), a calorimeter, as well as a barrel and two endcap toroidal
magnets supporting the muon spectrometer.  The ID consists of silicon
pixel, silicon microstrip, and transition radiation detectors and
provides precision tracking of charged particles for pseudorapidity
$\left|\eta\right| < 2.5$~\cite{fn1}.
The calorimeter, placed outside
the solenoid, covers $\left|\eta\right| < 4.9$ and is composed of
sampling electromagnetic and hadronic calorimeters with either liquid
argon or scintillating tiles as the active media. The muon
spectrometer surrounds the calorimeters and consists of a system of
precision tracking chambers in $|\eta|<2.7$, and detectors for
triggering in $|\eta|<2.4$.

The analysis is based on data recorded by the ATLAS detector in 2011
corresponding to 4.7~\ifb\ of integrated luminosity with the LHC
operating at a $pp$ center-of-mass energy of $7$\,\TeV.  
The data were collected requiring either a single-lepton (electron or muon) or an \met\ trigger.
The combined trigger efficiency is $>98\%$ for the chosen selection criteria on leptons and \met.
%The data were
%collected using a logical OR of a \met\ and a single-lepton (electron
%or muon) trigger.  The combined triggers reach plateau efficiencies
%$>$98\% for the chosen selection criteria on \met\ and the leptons.
Requirements that ensure the quality of beam conditions, detector
performance and data are imposed.

Monte Carlo (MC) event samples using the full ATLAS detector
simulation~\cite{:2010wqa} based on the {\tt Geant4}
program~\cite{Agostinelli:2002hh} 
are used to aid in the description of the background and to model the SUSY signal.
The effect of multiple $pp$ interactions per bunch crossing is also
simulated~\cite{ATLAS-CONF-2012-064}.  Production of top quark pairs is simulated with
\mcatnlov~\cite{Frixione:2005vw,Frixione:2008yi},
alternatively using \alpgenv~\cite{Mangano:2002ea} and
\powhegv~\cite{Nason:2004rx,Frixione:2007vw,Alioli:2010xd}.  The data
modeling is improved for high jet multiplicities by reweighting the
\mcatnlo\ sample to match the jet multiplicity distribution in
\alpgen.  
Uncertainties associated with initial- and final-state radiation (ISR/FSR)~\cite{ATLAS:2012al} 
are assessed using \acermcv~\cite{Kersevan:2004yg} samples.
A top quark mass of $172.5$\,\GeV\ is used consistently.  $W$ and
$Z$/$\gamma^*$ production in association with jets are each modeled with
\alpgen. Diboson $VV$ ($WW$, $WZ$ and $ZZ$) production is simulated with
\alpgen\ and cross-checked with
\herwigv~\cite{Corcella:2000bw}. Single top production is modeled with
\mcatnlo, and \ttbar\ events produced in association with $Z$, $W$ or
$WW$ ($\ttbar+V$) are generated with \madgraphv~\cite{Alwall:2007st}.
Next-to-leading-order (NLO) PDFs {\tt CT10}~\cite{Lai:2010vv} are used
with all NLO MC samples.  
For all other samples, LO PDFs are used: 
 {\tt MRSTmcal}~\cite{Martin:2009iq} with \herwig,
and \cteqVER{6L1}~\cite{Pumplin:2002vw} with \alpgen\ and \madgraph. 
Fragmentation and hadronization for the
\alpgen\ and \mcatnlo\ samples are performed with \herwig, using
\jimmyv~\cite{Butterworth:1996zw} for the underlying event, and for
the \madgraph\ samples \pythiav~\cite{Sjostrand:2006za} is used.  The
\ttbar, single top and $\ttbar+V$ production cross sections are
normalized to approximate next-to-next-to-leading order
(NNLO)~\cite{Aliev:2010zk}, next-to-next-to-leading-logarithmic
accuracy
(NLO+NNLL)~\cite{Kidonakis:2011wy,Kidonakis:2010ux,Kidonakis:2010tc}
and NLO~\cite{Campbell:2012dh} calculations, respectively.  QCD NNLO
\fewz~\cite{Gavin:2010az} inclusive $W$ and $Z$ cross sections are
used for the normalization of the $W$+jets and $Z$+jets
processes. Expected diboson yields are normalized using NLO QCD
predictions obtained with
\mcfm~\cite{Campbell:1999ah,Campbell:2011bn}.

Stop pair production is modeled using
\herwigppv~\cite{Bahr:2008pv}. 
The $\tone$ is chosen to be mostly the partner 
of the right-handed top quark, and the \ninoone\ to be almost a pure bino.
A signal grid is generated with a step
size of $50$\,\GeV\ both for the stop and LSP mass values.  
Signal cross sections are calculated to 
NLO in the strong coupling constant, including the resummation of soft
gluon emission at next-to-leading-logarithmic accuracy
(NLO+NLL)~\cite{Beenakker:1997ut,Beenakker:2010nq,Beenakker:2011fu}.
The nominal cross section and the uncertainty are taken from an
envelope of cross section predictions using different PDF sets and
factorization and renormalization scales~\cite{Kramer:2012bx}.
The $\tone\tone$  cross section for $m_{\tone}  = 400\,\GeV$ is $(0. 21 \pm 0.03)$\,pb.

Events must pass basic quality criteria to reject detector noise and
non-collision backgrounds~\cite{ATLAS-CONF-2012-020,Aad:2010ad} and
are required to have $\ge 1$ reconstructed primary vertex associated
with five or more tracks with transverse momentum $\pt > 0.4$\,\GeV.
Events are retained if
they contain exactly one muon~\cite{muons} with $\left|\eta\right| <
2.4$ and $\pt > 20$\,\GeV\ or one electron passing
`tight'~\cite{Aad:2011mk} selection criteria with $\left|\eta\right| <
2.47$ and $\pt > 25$\,\GeV.  Leptons are required to be isolated from
other particles.  The scalar sum of the transverse momenta of tracks
above $1$\,\GeV\ within a cone of size $\mathrm{\Delta} R = 0.2$ around
the lepton candidate is required to be $<$~10\% of the electron \pt,
and $<1.8$\,\GeV\ for the muon.  Events are rejected if
they contain additional leptons passing looser selection
criteria~\cite{atlas1lepFull2011Public}.  Jets are reconstructed from
three-dimensional calorimeter energy clusters using the anti-$k_t$ jet
clustering algorithm~\cite{Cacciari:2008gp} with a radius parameter of
$0.4$.  The jet energy is corrected for the effects of calorimeter
non-compensation and inhomogeneities using \pt- and
$\eta$-dependent calibration factors based on MC simulations and
validated with extensive test-beam and collision-data
studies~\cite{Aad:2011he}.  To suppress jet background originating
from uncorrelated soft collisions, $\ge 75$\% of the summed \pT\ of
all tracks associated to a jet must come from tracks associated to the
selected primary vertex.  Events with four or more jets 
with $\left|\eta\right| < 2.5$ and $\pt >$ 80, 60, 40, and 25\,\GeV\ are selected.  
At least one jet needs to be identified as a $b$-jet, 
which is a jet containing a $b$-hadron decay.
These are identified using the
`MV1' $b$-tagging algorithm~\cite{ATLAS-CONF-2012-043} 
which exploits both impact parameter and secondary vertex information.
An operating point is employed corresponding to an average 75\% $b$-tagging
efficiency and to a $<2$\% misidentification rate for light-quark/gluon
jets for jets with $\pt > 20$\,\GeV\ and $\left|\eta\right| < 2.5$ in
\ttbar\ MC events.

Ambiguities between overlapping leptons and jets are
resolved by discarding
either the jet or lepton candidates~\cite{atlas1lepFull2011Public} based
on their separation $\mathrm{\Delta} R$.  The measurement of \met\ is
based on the transverse momenta of all electron and muon candidates,
all jets after overlap removal, and all calorimeter energy clusters
not associated to such objects.  The background is reduced by
requiring $\mathrm{\Delta}\phi_{\mathrm{min}}>0.8$, where
$\mathrm{\Delta}\phi_{\mathrm{min}}$ is the minimum
azimuthal separation between the two highest \pt\ jets and the missing
transverse momentum direction.  
A requirement on the three-jet mass $m_{jjj}$ of the hadronically decaying top quark specifically rejects the
dileptonic \ttbar\ background, where both $W$
bosons from the top quarks decay leptonically.  The jet-jet pair
having invariant mass $>60$\,\GeV\ and the smallest
$\mathrm{\Delta}R$ is selected to form the hadronically decaying $W$ boson. The
mass $m_{jjj}$ is reconstructed including a third jet closest in
$\mathrm{\Delta}R$ to the hadronic $W$ boson momentum vector and
$130$\,\GeV\ $< m_{jjj} < 205$\,\GeV\ is required.

Five signal regions (SR A--E) are defined in order to optimize the sensitivity
for different stop and LSP masses.  For increasing stop mass and
increasing mass difference between stop and LSP the requirements are
tightened on \met, on the ratio $\metsig$, where $\HT$ is the scalar
sum of the momenta of the four selected jets with highest \pt,
and on the transverse mass \mt~\cite{fn3}, as shown in Table~\ref{tab:sr}.
The number of observed events in each SR after applying all selection criteria
are given in Table~\ref{tab:fitresults}.
\begin{table}[htb!]
\begin{center}
\caption{\label{tab:sr} Selection requirements defining the SR A--E.}
\begin{tabular*}{\columnwidth}{lccccc}
\hline
Requirement          & SR A  & SR B & SR C & SR D & SR E \\
\hline
\met\ [\GeV]  $>$         & 150   & 150  & 150  & 225  & 275 \\
\metsig\ [\GeV$^{1/2}$] $>$ & 7     & 9    & 11   & 11   & 11  \\
\mt\ [\GeV]    $>$         & 120   & 120  & 120  & 130  & 140 \\
\hline
\end{tabular*}
\end{center}
\end{table}

The product of the kinematic acceptance, detector and reconstruction efficiency
($A\cdot\epsilon$) varies between 4\% and 0.3\% for SR A and between
3\% and 0.01\% for SR E as the stop-LSP mass difference varies between
$550$\,\GeV\ and $250$\,\GeV.

The dominant background arises from dileptonic \ttbar\ events in which
one of the leptons is either not identified, is outside the detector
acceptance, or is a hadronically decaying $\tau$ lepton.  In all these
cases, the \ttbar\ decay products include two or more high-\pt\
neutrinos, resulting in large \met\ and \mt.  
Three control
regions (CRs) enriched in dileptonic \ttbar\ events ({\em 2-lep TR}),
single-leptonic $\ttbar$ events ({\em 1-lep TR}), and \Wjets\ events
({\em 1-lep WR}) are designed
to normalize the corresponding backgrounds using data.
The {\em 2-lep TR} differs from the SRs by selecting events
with exactly two leptons, applying no requirements on \mt, \metsig\
and $m_{jjj}$, and by requiring $\met > 125$\,\GeV.  The {\em 1-lep TR}
and {\em 1-lep WR} have selection criteria identical to SR A, 
except the \mt\ requirement is changed to $60 < \mt < 90$\,\GeV\ and the {\em 1-lep WR} has a \bjet\ veto instead of a
$b$-jet requirement.
\ttbar\ production accounts for $>90$\% of events in
the top CRs and $W$+jets production for $>60$\% in the $W$ CR. 
The signal contamination reaches a maximum of 8\% in the {\em 2-lep TR} for $m_{\tone}$ = 200\,\GeV.
The multijet background, which mainly originates from jets misidentified as leptons,
is estimated using the matrix method~\cite{atlas1lepFull2011Public}.
Other background contributions
($VV$, $\ttbar+V$ and single top) are estimated using MC
simulation normalized to the theoretical cross sections.  The \Zjets\ background is found to be negligible.

Good agreement is observed between data and the SM prediction before using the
CRs to normalize the \ttbar\ and $W$+jets backgrounds. As an
example, Fig.~\ref{fig:controlDistributions1} shows the agreement of the
\met\ distributions in the {\em 2-lep TR}, and the \met\ distribution in SR
A. In addition, the \mt\ distribution for a looser requirements region
 --- $\met > 40$\,\GeV\ and no requirements on \metsig\ and $m_{jjj}$ (pre-selection) --- is shown.

\begin{figure}
  \centering
  \includegraphics[width=0.9\columnwidth]{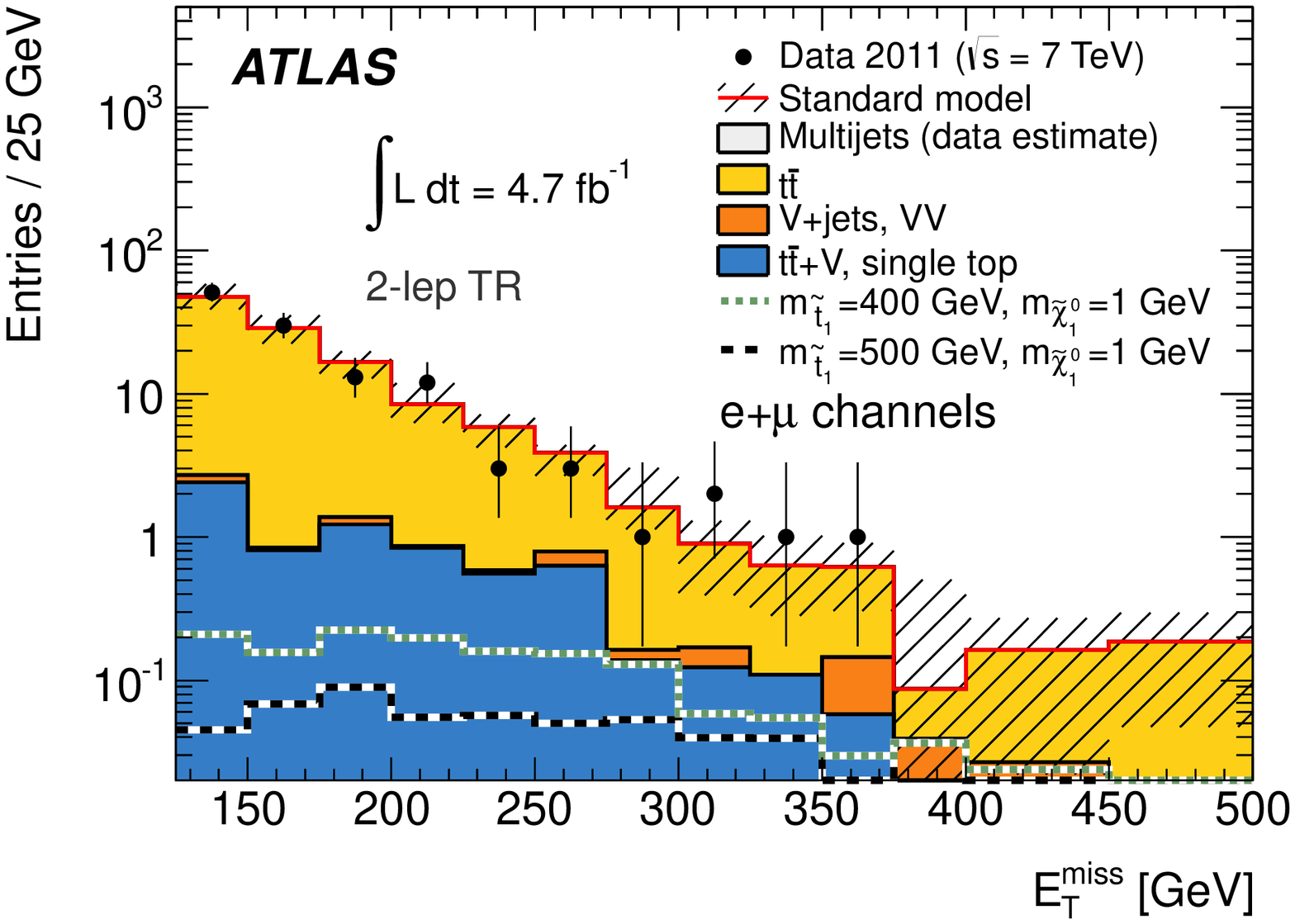}
  \includegraphics[width=0.9\columnwidth]{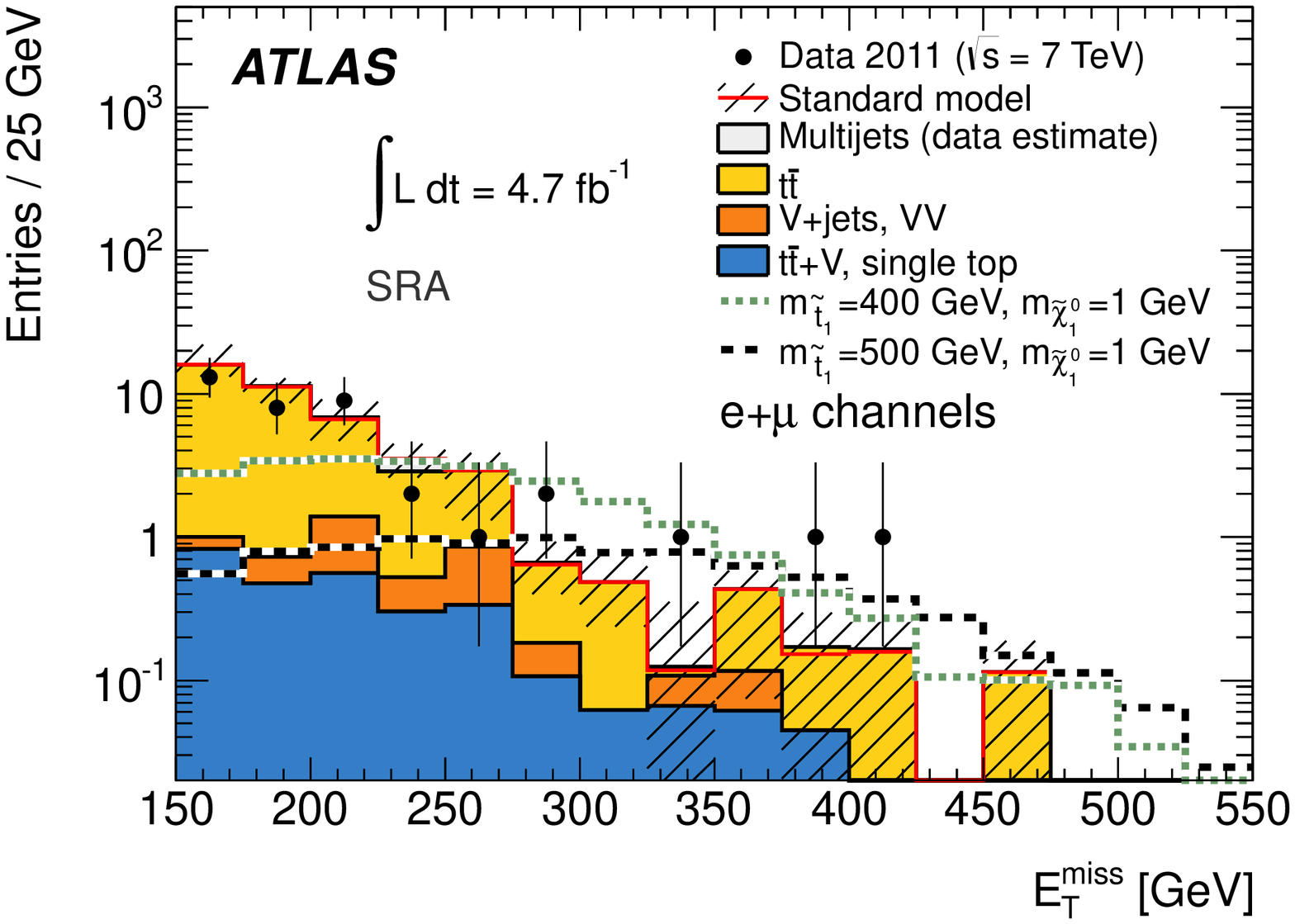}
    \includegraphics[width=0.9\columnwidth]{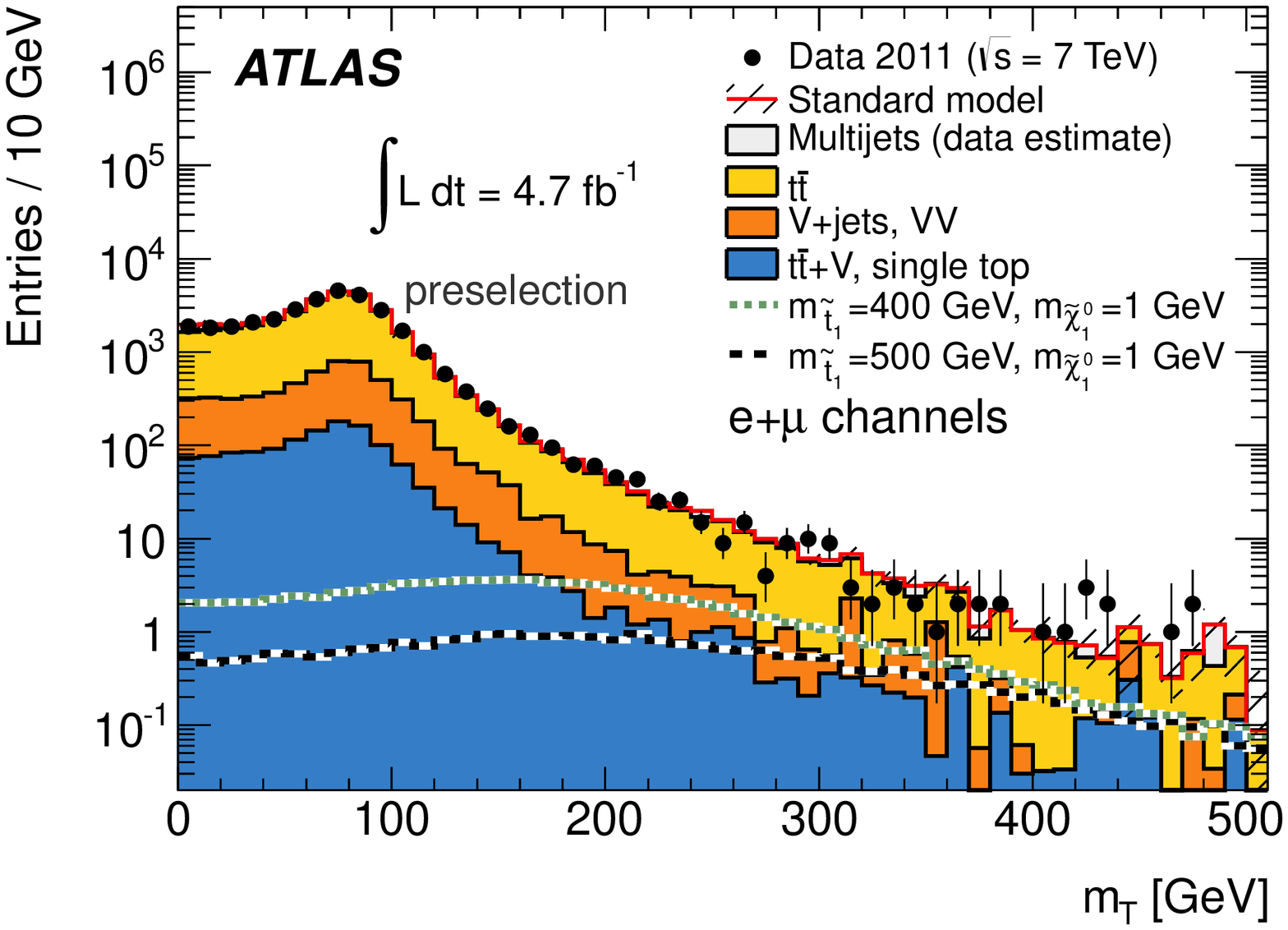}
  \caption{
     Top: \met\ distributions for the {\em 2-lep TR}.
     Center: \met\ distributions for SR A.
     Bottom:  \mt\ distribution for the pre-selection requirements (see text).
     All plots show the electron and muon combined channel, before
     normalization fits. 
    Hatched areas indicate the combined uncertainty due to MC sample size and the jet energy scale.
  }
  \label{fig:controlDistributions1}
\end{figure}

Simultaneous fits to the numbers of observed events in the three CRs
and one SR at a time are performed to normalize the \ttbar\ and
$W$+jets background estimates as well as to search for an
excess from a potential signal contribution.  The {\em 1-lep} and {\em 2-lep TRs}
have \ttbar\ normalizations that float independently and that are
found to be in good agreement with each other.  The \ttbar\ estimates
in the SRs are based on the {\em 2-lep TR}, as this minimizes the
extrapolation uncertainties in the fit.  Systematic uncertainties
are treated as nuisance parameters with Gaussian probability density
functions.

The dominant systematic uncertainties in the fitted \ttbar\ background
estimate are theoretical and modeling uncertainties,
which affect the event kinematics and thus the extrapolation from the CR to the various SRs.
They are
determined by using different generators (\mcatnlo, \powheg\ and
\alpgen), different showering models (\herwig\ and \pythia) and by
varying ISR/FSR parameters, and amount to 10--30\%.
% on the extrapolation from the CR to the various SRs. 
Electroweak single top
production is associated with an $8\%$ theoretical
uncertainty~\cite{Kidonakis:2011wy,Kidonakis:2010ux,Kidonakis:2010tc}
and the $\ttbar+V$ background has a $30\%$
uncertainty~\cite{Campbell:2012dh}.  The difference between \alpgen\
and \herwig\ predictions is used to assess the uncertainty on the
diboson background, and the uncertainty on the multijet background is
based on the matrix method.  Both of these uncertainties are estimated
as $100\%$.

Experimental uncertainties affect the signal and background yields,
including those normalized in CRs.
They are estimated by aid of MC events and are 
dominated by uncertainties in the jet energy scale,
jet energy resolution and $b$-tagging.  Uncertainties
related to the trigger and lepton reconstruction and identification
(momentum and energy scales, resolutions and efficiencies) give
smaller contributions.  Other small uncertainties are due to modeling
of multiple $pp$ interactions, the integrated 
luminosity, and the limited numbers of MC and data events.  The
uncertainty on $A\cdot\epsilon$ varies between $9$\% and $16$\% as the
simulated stop-LSP mass difference varies between $550$\,\GeV\ (SR E) and
$250$\,\GeV\ (SR A and B).

\begin{table*}[!htbp]
\begin{center}
\caption{Numbers of observed events in the five signal regions and three background control 
regions, as well as their estimated values and all (statistical and systematic) uncertainties from a fit to the control regions only, for
the electron and muon combined channel. 
The expected numbers of signal events for
$m_{\tone}=400\,\GeV$ ($500\,\GeV$) and $m_{\tilde{\chi}^{0}_{1}}=1\,\GeV$
for benchmark points 1 (2) are listed for comparison.
The central values of
the fitted sum of backgrounds in the control regions agree with the
observations by construction.  
Furthermore, $p_0$-values and 95\% \CLs\ observed (expected) upper limits on beyond-SM events
are given, using simultaneous fits including one SR at a time and all CRs.
}
\label{tab:fitresults}
\setlength{\tabcolsep}{0.0pc}
{\footnotesize
\begin{tabular*}{\textwidth}{@{\extracolsep{\fill}}lcccccccc}%rrrrr|rrr}
  \noalign{\smallskip}\hline\noalign{\smallskip}
  {\bf Regions}        &  SR A & SR B & SR C & SR D  & SR E &  {\em 2-lep TR} & {\em 1-lep TR}  & {\em 1-lep WR}   \\[-0.05cm]
  \noalign{\smallskip}\hline\noalign{\smallskip}
  \ttbar\  & $36 \pm 5$          & $27 \pm 4$ & $11 \pm 2$  & $4.9 \pm 1.3$  & $1.3 \pm 0.6$  & $109 \pm 10$  & $364 \pm 23$  & $59 \pm 19$                \\
  $\ttbar+V$, single top & $2.9 \pm 0.7$  & $2.5 \pm 0.6$   & $1.6 \pm 0.3$   & $0.9 \pm 0.3$   & $0.4 \pm 0.1$  & $7.2 \pm 1.3$  & $18 \pm 3$   & $6.1 \pm 1.6$                \\
  $V$+jets, $VV$   & $2.5 \pm 1.3$    & $1.7 \pm 0.8$   & $0.4 \pm  0.1$  & $0.3 \pm 0.1$  & $0.1 \pm 0.1$  & $1.6 \pm 0.8$  & $38 \pm 11$ & $162 \pm 23$                \\
Multijet  & $0.4 \pm 0.4$ & $0.3 \pm 0.3$& $0.3 \pm 0.3$ & $0.3 \pm 0.3$ &$0.0 {}^{+0.3}_{-0.0}$ & $0.0 {}^{+0.6}_{-0.0}$ & $1.7 \pm 1.7$ & $0.8 \pm 0.8$\\
  \noalign{\smallskip}\hline\noalign{\smallskip}
  Total background & $42 \pm 6$   & $31 \pm 4$  &  $13 \pm 2$  &   $6.4 \pm 1.4$  & $1.8 \pm 0.7$  & $118 \pm 10$  & $421 \pm 20$  & $228 \pm 15$              \\
 Signal benchmark 1 (2)  &   $25.6$ ($8.8$) & $23.0$ ($8.1$) & $17.5$ ($6.9$) & $13.5$ ($6.2$) & $7.1$ ($4.5$) & $1.7$ ($0.6$) & $2.3$ ($0.6$) & $0.4$ ($0.1$) \\
  Observed events &   $38$ & $25$ & $15$ & $8$ & $5$ &  $118$ & $421$ & $228$\\
  \noalign{\smallskip}\hline
  \noalign{\smallskip}\hline\noalign{\smallskip}
  $p_0$-values &   $0.50$ & $0.50$ & $0.32$ & $0.24$ & $0.015$ & - & - & -\\
  \noalign{\smallskip}\hline\noalign{\smallskip}
   Obs. (exp.) $N_{\mathrm{BSM}}<$ & $15.1$ ($17.2$) & $10.1$ ($13.8$) &  $10.8$ ($9.2$) &  $8.4$ ($7.0$) & $8.2$ ($4.6$) & - & - & -\\
  \noalign{\smallskip}\hline\noalign{\smallskip}
\end{tabular*}}
\end{center}
\end{table*}

Table~\ref{tab:fitresults} shows the results of the background fit to the CRs,
extrapolated to the SRs.
The fitted numbers of \ttbar\ and $W$+jets events are compatible with
the MC predictions, with factors of 1.01 and 0.90 applied,
respectively.
To assess the agreement between the SM expectation and the observation in the SRs,
a second set of simultaneous fits including one SR at a time and all CRs is
performed.
The $p_0$-values (probing the background-only hypothesis) obtained are given in Table~\ref{tab:fitresults}. 
No significant excess of events is found.

One-sided exclusion limits are derived using
the \CLs\ method~\cite{clsread}, based on the same simultaneous fit method
but taking the predicted signal contamination in the CRs into account.
To obtain the best expected combined exclusion limit, a mapping
in the stop-LSP mass plane is constructed by selecting the SR
with the lowest expected \CLs\ value for each grid point. 
The expected and observed 95\% \CLs\ 
exclusion limits are displayed in Fig.~\ref{fig:stopExclusion}.  
Stop masses are
excluded between $230$\,\GeV\ and $440$\,\GeV\ for massless LSPs, and stop masses
around $400$\,\GeV\ are excluded for LSP masses up to $125$\,\GeV.
These values are derived from the $-1\,\sigma_{\mathrm{theory}}^{\mathrm{SUSY}}$  observed limit contour.
These stop mass limits significantly extend previous results~\cite{PhysRevLett.104.251801,D0_stopSearch}.
\begin{figure}
  \centering
\includegraphics[width=\columnwidth]{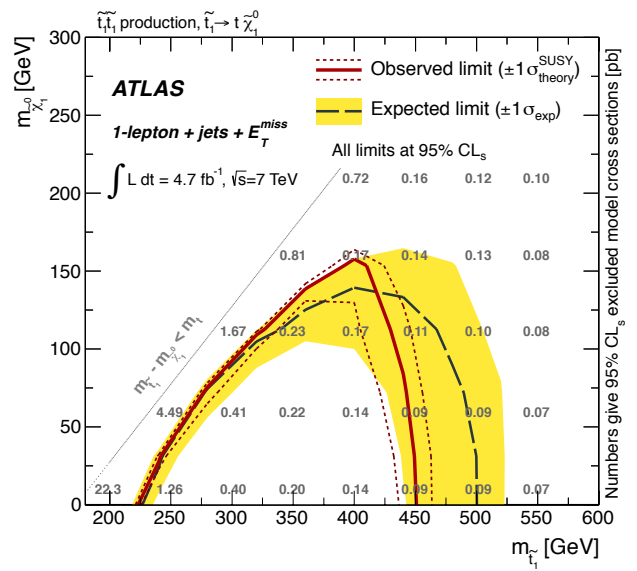}
  \caption{
    Expected (dashed) and observed (solid curve) 95\% \CLs\ excluded region (under the curve) in the plane
  of $m_{\ninoone}$ vs. $m_{\stop_1}$, assuming $BR(\stop_1 \to t \ninoone) = 100\%$.
All uncertainties except the
%theoretical
 signal cross-section uncertainties are included. The contours of the shaded band
around the expected limit are the $\pm 1\,\sigma$ results.
The dotted lines around the observed limit illustrate the change in the
observed limit as the nominal signal cross section is scaled up and
down by the theoretical uncertainty.  
The overlaid numbers give the 95\% \CLs\ upper limit on the signal cross
section, in pb.  
  }
  \label{fig:stopExclusion}
\end{figure}
Limits on beyond-SM contributions are derived from the same simultaneous fit
but without signal model-dependent inputs (i.e. without signal contamination in the CRs, 
and without signal systematic uncertainties).
The resulting limits are shown at the bottom of Table~\ref{tab:fitresults}. 

In summary, a search for stop pair production is presented in final
states with one isolated lepton, jets, and missing transverse momentum
in $\sqrt{s}=7$\,\TeV\ $pp$ collisions corresponding to $4.7$~\ifb\ of
ATLAS 2011 data.  
Each stop is assumed to decay to a top quark and a
long-lived undetected neutral particle.  No significant excess of
events above the rate predicted by the standard model is observed and
95\% \CLs\ upper limits are set on the stop mass in the stop-LSP mass
plane, significantly extending previous stop-mass limits.

% Acknowledgements for papers with collision data
% Version 9-Aug-2012

\section{Acknowledgements}

We thank CERN for the very successful operation of the LHC, as well as the
support staff from our institutions without whom ATLAS could not be
operated efficiently.
We acknowledge the support of ANPCyT, Argentina; YerPhI,
Armenia; ARC, Australia; BMWF, Austria; ANAS, Azerbaijan; SSTC, Belarus; CNPq and FAPESP,
Brazil; NSERC, NRC and CFI, Canada; CERN; CONICYT, Chile; CAS, MOST and NSFC,
China; COLCIENCIAS, Colombia; MSMT CR, MPO CR and VSC CR, Czech Republic;
DNRF, DNSRC and Lundbeck Foundation, Denmark; EPLANET and ERC, European Union;
IN2P3-CNRS, CEA-DSM/IRFU, France; GNSF, Georgia; BMBF, DFG, HGF, MPG and AvH
Foundation, Germany; GSRT, Greece; ISF, MINERVA, GIF, DIP and Benoziyo Center,
Israel; INFN, Italy; MEXT and JSPS, Japan; CNRST, Morocco; FOM and NWO,
Netherlands; RCN, Norway; MNiSW, Poland; GRICES and FCT, Portugal; MERYS
(MECTS), Romania; MES of Russia and ROSATOM, Russian Federation; JINR; MSTD,
Serbia; MSSR, Slovakia; ARRS and MVZT, Slovenia; DST/NRF, South Africa;
MICINN, Spain; SRC and Wallenberg Foundation, Sweden; SER, SNSF and Cantons of
Bern and Geneva, Switzerland; NSC, Taiwan; TAEK, Turkey; STFC, the Royal
Society and Leverhulme Trust, United Kingdom; DOE and NSF, United States of
America.
The crucial computing support from all WLCG partners is acknowledged
gratefully, in particular from CERN and the ATLAS Tier-1 facilities at
TRIUMF (Canada), NDGF (Denmark, Norway, Sweden), CC-IN2P3 (France),
KIT/GridKA (Germany), INFN-CNAF (Italy), NL-T1 (Netherlands), PIC (Spain),
ASGC (Taiwan), RAL (UK) and BNL (USA) and in the Tier-2 facilities
worldwide.

\bibliography{PUB_topLSP1lepton}
\onecolumngrid
\clearpage
\input{atlas_authlist.tex} 

\end{document}

%% file: atlas_authlist.tex
% ATLAS Collaboration author list
% Data extracted on 11-Jul-2012 for paper reference SUSY-2012-06
%\documentclass[11pt]{article}
%\usepackage{a4wide}\begin{document}
\begin{flushleft}
{\Large The ATLAS Collaboration}

\bigskip

G.~Aad$^{\rm 48}$,
T.~Abajyan$^{\rm 21}$,
B.~Abbott$^{\rm 111}$,
J.~Abdallah$^{\rm 12}$,
S.~Abdel~Khalek$^{\rm 115}$,
A.A.~Abdelalim$^{\rm 49}$,
O.~Abdinov$^{\rm 11}$,
R.~Aben$^{\rm 105}$,
B.~Abi$^{\rm 112}$,
M.~Abolins$^{\rm 88}$
% ...
\bigskip

% ...
$^{11}$ Institute of Physics, Azerbaijan Academy of Sciences, Baku, Azerbaijan\\
$^{12}$ Institut de F\'{i}sica d'Altes Energies and Departament de F\'{i}sica de la Universitat Aut\`{o}noma de Barcelona and ICREA, Barcelona, Spain\\
% ...
$^{21}$ Physikalisches Institut, University of Bonn, Bonn, Germany\\
% ...
$^{48}$ Fakult\"{a}t f\"{u}r Mathematik und Physik, Albert-Ludwigs-Universit\"{a}t, Freiburg, Germany\\
$^{49}$ Section de Physique, Universit\'{e} de Gen\`{e}ve, Geneva, Switzerland\\
% ...
$^{88}$ Department of Physics and Astronomy, Michigan State University, East Lansing MI, United States of America\\
% ...
$^{105}$ Nikhef National Institute for Subatomic Physics and University of Amsterdam, Amsterdam, Netherlands\\
% ...
$^{111}$ Homer L. Dodge Department of Physics and Astronomy, University of Oklahoma, Norman OK, United States of America\\
$^{112}$ Department of Physics, Oklahoma State University, Stillwater OK, United States of America\\
% ...
$^{115}$ LAL, Universit\'{e} Paris-Sud and CNRS/IN2P3, Orsay, France\\
% ...
\end{flushleft}

%\end{document}
% Created with TexConverter

%% file: PUB_topLSP1lepton.bbl
\begin{thebibliography}{66}
\bibitem [{fn1()}]{fn1}%
  \BibitemOpen
  \href@noop {} {}\bibinfo {howpublished} {ATLAS uses a right-handed coordinate
  system with its origin at the nominal interaction point in the centre of the
  detector and the $z$-axis along the beam pipe. Cylindrical coordinates
  $(r,\phi)$ are used in the transverse plane, $\phi$ being the azimuthal angle
  around the beam pipe. The pseudorapidity $\eta$ is defined in terms of the
  polar angle $\theta$ by $\eta=-\ln\tan(\theta/2)$, and $\mathrm{\Delta} R =
  \sqrt{(\mathrm{\Delta} \eta)^2 + (\mathrm{\Delta} \phi)^2}$}\BibitemShut
  {NoStop}%
\bibitem [{fn3()}]{fn3}%
  \BibitemOpen
  \href@noop {} {}\bibinfo {howpublished} {\mt is defined as $\mt^2 = 2
  \pT^{\mathrm{lep}} \met (1 - \cos( \mathrm{\Delta} \phi))$, where
  $\mathrm{\Delta} \phi$ is the azimuthal angle between the lepton and missing
  momentum direction.}\BibitemShut {Stop}%
\end{thebibliography}
